\documentclass{lg}
\usepackage{times}
\usepackage{graphicx} 
\def\la{\hbox{{\lower -2.5pt\hbox{$<$}}\hskip -8pt\raise
-2.5pt\hbox{$\sim$}}}
\def\ga{\hbox{{\lower -2.5pt\hbox{$>$}}\hskip -8pt\raise
-2.5pt\hbox{$\sim$}}}

\begin{document}

\title{Particle Dark Matter Constraints from the Draco Dwarf Galaxy}

\author[1]{C. TYLER}
\affil[1]{Department of Astronomy \& Astrophysics, \\ The
University of Chicago, Chicago, IL 60637}




\pagestyle{plain}

\pagenumbering{arabic}

\titleheight{4cm} 
\maketitle

\begin{abstract}
It is widely thought that neutralinos, the lightest
supersymmetric particles, could comprise most of the dark
matter.  If so, then dark halos will emit radio and gamma
ray signals initiated by neutralino annihilation.
A particularly promising place to look for these
indicators is at the center of the local group dwarf
spheroidal galaxy Draco, and recent measurements of
the motion of its stars have revealed it to be
an even better target for dark matter detection
than previously thought.  We compute limits on WIMP
properties for various models of Draco's dark matter
halo.  We find that if the halo is nearly isothermal,
as the new measurements indicate, then current gamma ray flux
limits prohibit much of the neutralino parameter space.
If Draco has a moderate magnetic field, then current radio
limits can rule out more of it.  These results
are appreciably stronger than other current
constraints, and so acquiring more detailed data on
Draco's density profile becomes one of the most
promising avenues for identifying dark matter.
\end{abstract}

\section{Introduction}

Despite the popularity of the neutralino as a dark matter candidate,
efforts to constrain its properties have only been
able to rule out a small fraction of the relevant parameter space.
Direct detection experiments based on scattering with nuclei, and
indirect searches based on neutralino annihilation, have shared this
difficulty.  But recent results can improve the situation.
New observations of the Draco dwarf galaxy reveal that
it is strongly dark matter dominated, and that its dark matter
distribution is at least nearly isothermal.  In this paper we
investigate the detectability of WIMP annihilation signals from
several appropriate halo models.  Interestingly, if Draco is indeed
isothermal down to very small radii, we will show that current
gamma ray measurements rule out a significant fraction of the
neutralino parameter space.  If one makes an additional
assumption that Draco harbors magnetic fields, then current
radio measurements rule out additional parameter space.
For example, a 1~$\mu {\rm G}$ field would eliminate most of it.
Other halo models yield weaker current limits, some of which
become significant with upcoming detectors.

The choice of a particular dark matter distribution is
critical to both gamma ray and radio limits, and we devote section
2 of this paper to that consideration.  The assumption of a magnetic
field in Draco is discussed in section 4.  The remainder of this
introductory section provides the relevant background.

The cold dark matter (CDM) scenario starts with a weakly interacting
massive particle (WIMP), created abundantly in the big bang and
surviving today to dominate the matter density in galaxies.  The
velocities of stellar orbits imply that dark matter occupies
galaxies with a particular radial density profile.  Over a wide
range in radius, the density appears to fall off as
$\rho \sim r^{-2}$, which is the same structure as would be found
in a self-gravitating isothermal sphere.  But when simulations of
the cosmological CDM evolution are performed, taking into account the
influence of {\it N}-body dynamics and tidal interactions, the profiles
are found to be more complicated at large and small $r$
(e.g.,~\citet{nfw},~\citet{moore99}).  We discuss these profiles
in more detail in subsection 2.1.

Simulations universally generate a central cusp in every dark halo.
Slopes vary, but $\rho \sim r^{-\gamma}$ remains for some positive
$\gamma$, and therefore the density blows up at the center.
However, a controversy ensues at this point because observations
often differ.  21~cm rotation curves from low surface brightness
(LSB) galaxies have presented evidence for central cores
of approximately constant density~\citep{mb88}.
A separate study of 16 LSB galaxies contends
that beam smearing effects render these measurements unreliable,
such that cusps are still fully consistent with the
data~\citep{brdb00}.  Many of the same galaxies have since been
re-examined with high resolution optical rotation curves,
and that work found 30 LSB galaxies evidently
possessing cores, not CDM cusps~\citep{bmr01}.
Dwarf galaxies too have long been considered problematic
for CDM because they typically have the linearly rising rotation
curves of a constant density core~\citep{nfw}, but a subsequent
study of 20 dwarf galaxies gives rise to interpretations ranging
from cores to steep $r^{-2}$ cusps~\citep{bs01}.  These issues
constitute an area of much debate, but in general, because
many observations favor cores over cusps, it is of great interest
to determine what processes might prevent or destroy a cusp.

If a WIMP cusp were to realize, and if WIMPs exist today in equal
numbers with antiWIMPs (or if they are their own antiparticles),
then we can expect densities high enough for dark matter
annihilation to become significant.  If the annihilation products
lead to observable particles, then cusps become hot spots
to search for WIMPs.

Perhaps the most likely WIMP candidate is the neutralino ($\chi$),
which is the lightest supersymmetric particle (LSP) in the minimal
supersymmetric extension of the standard model.  Supersymmetry
is a symmetry relating fermions to bosons, and thus it
introduces a new class of particles which are ``superpartners''
to standard model particles.  The theory helps to remedy
the hierarchy problem set by the electroweak and Planck scales.
In models with conservation of R-parity, a supersymmetric particle
is prohibited from decaying into only non-supersymmetric
ones, and since the neutralino is the LSP, there is no
supersymmetric state of lower energy available.
Thus the $\chi$ is stable, making it a well motivated dark
matter candidate (for a review of supersymmetric dark matter,
see~\citet{jkg96}).

Each $\chi \bar{\chi}$ annihilation produces
quarks which bind primarily into pions.  Neutral pions usually
decay to gamma rays; charged pions usually decay to neutrinos and
muons, and subsequently electrons (and positrons), which generate radio
waves by synchrotron radiation in an ambient magnetic field.
(This approach does not wholly rely on supersymmetry, but rather
on some DM particle which annihilates into pions.)
Therefore, a cusp produces observable radio
and gamma radiation~\citep{bgz92, bbm94}; the trick
is just finding a galaxy with a cusp.  

The candidate galaxy should have a very high mass-to-light ratio
($M/L$) to ensure that its internal gravitation is DM dominated,
and it should be nearby to facilitate the detection of a weak signal.
As it turns out, the highest known $M/L$ comes from a dwarf satellite
of our own Galaxy.  The Draco dwarf spheroidal galaxy, only 79~kpc
distant from Earth~\citep{vdb00}, was recently re-examined for detailed
stellar velocities~\citep{kweg01}.  A new, higher than previously
estimated $M/L$ was found to be $440 \pm 240M_{\odot}/L_{\odot}$,
with a mass of $M \simeq 8.6 \times 10^7~M_{\odot}$.
Draco has the mass of a dwarf galaxy and the starlight of a globular
cluster.

Furthermore, the candidate galaxy should have a high central WIMP
density, without contaminants such as massive baryonic clouds
or a central black hole.  The stellar velocity
dispersion measurements in Draco are consistent with an isothermal
dark matter halo~\citep{kweg01}, implying a steep central density.
Its present hydrogen gas population is less than
$450~M_{\odot}$~\citep{young99}, with no evidence of a central
black hole.  The Sloan Digital Sky Survey reported that
tidal deformation can only exist at a level of $\leq 10^{-3}$ of
the central stellar surface density~\citep{sdss}, supporting the case
for dynamical relaxation and therefore meaningful extraction of
mass density from velocity information.  Given these properties,
we present particle dark matter constraints enabled by
observations of the stars in Draco.

The next section provides more detail
regarding halos with cusps and cores.  Section 3 follows with the
calculation of the observable flux created by WIMP annihilations,
including electron losses by inverse Compton scattering, which is
an important correction not usually considered in such
analyses.  Section 4 presents the results of these calculations,
using observational bounds from EGRET and the VLA, with a
discussion of anticipated improvements by new experiments, and
a comparison among various halo profiles.
We summarize in section 5.

\section{The Dark Halo}

A power law cusp's density profile has the form
$\rho \sim r^{-\gamma}$.  \citet{bt87} have put forth some
reasoning for the expectation of a singular isothermal sphere (SIS)
distribution to a first approximation, where $\gamma = 2$.  But do
steep cusps like this really exist in nature?

Observations of galaxy centers with the Hubble Space Telescope
(HST) shed some light on this question~\citep{c93,g96}.
These authors have constructed three dimensional luminosity
densities based on an assumed spherical geometry in the target
ellipticals and in the bulges of spirals, further assuming that the
light traces the mass in these systems.  Among power law
galaxy centers, two populations emerge:  bright galaxies (roughly
$M_V \leq -20$) with slopes in the range $0.2 \leq  \gamma
\leq 1.5$, and faint galaxies ($M_V \geq -20$) with $\gamma$
ranging between $1.5 \leq \gamma \leq 2.5$.  The peaks in this
bimodal distribution are at 0.8 and 1.9, respectively.
Many faint galaxies in the sample therefore have nearly
isothermal density profiles at their centers.

\subsection{Halo Models}

An ordinary kinetic-molecular gas, self-gravitating and held isothermal,
will assume the SIS density profile $\rho \sim r^{-2}$.  Instead, we
have a collisionless system of WIMP particles.  If they are
allowed to evolve under their collective gravitational
potential, the initial infall can be expected to be clumpy, and as a
result, the overall potential seen by any one WIMP is the superposition
of the central field and the temporary groupings.  Trajectories
are thus altered, and over a few orbital periods, a new smooth
centrally weighted distribution will be obtained.  This process was
termed ``violent relaxation'' by its author,~\citet{lb67}, wherein
the central collapse is achieved by transporting angular momentum
outward.

The solution at the end of this relaxation turns out to be the
Maxwellian velocity distribution with constant velocity dispersion
at every radius, although the population dwindles somewhat beyond the
virial radius, as those particles do not often interact with the lumpy
center and therefore do not join the infall.  One important caveat,
however, is that the violent relaxation process proceeds until
a steady state is achieved, whether or not that state is the
Maxwellian result of complete relaxation.  This means that it is
possible for the final state to retain features unique to its
own collapse sequence (one reason why simulations are useful
for this work).

Supposing the end state of the process is complete relaxation,
we have a system of self-gravitating particles
with a Maxwellian velocity distribution at every point
in space, the solution of which is again the SIS
density profile~\citep{bt87}.  If the system fails to achieve
complete relaxation, then the innermost orbits may not be fully
populated, and one might expect $\gamma < 2$ as is often seen in
numerical simulations.

Rotation curves of galaxies usually show a constant velocity
over a wide range of radial values.  On the contrary however,
some observations reveal core regions of about $1 - 10$~kpc in
radius which appear to enclose an $r$-independent
matter density.  This is evident in the rotation curves of spiral
galaxies which tend to rise linearly at small radii, and those of
some dwarf galaxies which rise linearly for the full extent of the
observed galaxy.  To model this, one typically invokes a
modified isothermal profile:  a spherically symmetric density
profile which smoothly blends a constant density core with
isothermal behavior outside the core.

In order to calculate observables from annihilation reactions,
one must start by assuming some profile.  For example, gamma rays
from annihilating neutralinos were originally calculated
by~\citet{bgz92} using $\rho \sim r^{-1.8}$ for the Galactic
center.  In~\citet{g93}, the same signature from the Large Magellanic
Cloud was calculated assuming $\rho \sim a^2/(r^2 + a^2)$, a modified
isothermal model with core radius $a$.  Non-constant central
densities with radial power law exponents have been employed
in different contexts:  $\gamma = 1$ (NFW:~\citet{nfw}, used
for example in~\citet{g00} and~\citet{bot02}), $\gamma = 1.5$
(Moore:~\citet{moore99},
used for example in~\citet{cm01}), and $\gamma = 2$
(isothermal: used for example in~\citet{bot02}).  It is worth
mentioning that $\gamma$ appears to vary with the resolution of the
CDM simulation, and at small radii, a realistic model of the high
cusp density can be expected to require more accurate simulations.

In this work, we consider seven distinct models of Draco.  These
include steep power-law cusps with $\gamma \leq 2$, the Moore profile,
and isothermal halos truncated by constant density cores.

A power law DM halo profile must include a small constant density
core inside the cusp for any $\gamma > 0$, due to the fact
that inside some radius, the annihilation rate gets so large
that the over-density is destroyed as fast as new infall
can fill the region.  The constant density region, found
by setting cusp forming time scale (the clump crossing time, or
approximately the free fall time $\sim 1 / \sqrt{G \bar{\rho}}$)
equal to the annihilation time scale ($1 / n_{\chi}
\langle \sigma v \rangle_{\chi \bar{\chi}}$),
establishes a characteristic inner radius~\citep{bgz92}.
For example, if $\gamma = 2$ (SIS), we have~\citep{bot02}:
\begin{equation}
R_{\rm min} = R_{\rm ext}\, \langle \sigma v
\rangle_{\chi \bar{\chi}}^{1 \over 2} \left[ {n_{\rm halo}
\over 4 \pi G m_{\chi} } \right]^{1 \over 4}\,,
\label{Rmin}
\end{equation}
where the $\chi \bar{\chi}$ annihilation rate (cross section $\times$
velocity) $\langle \sigma v \rangle_{\chi \bar{\chi}}$ is approximately
constant in the low velocity limit appropriate for galactic dark
matter~\citep{bgz92}, and the time scales mentioned above
are $\sim 5 \times 10^8$~years at $R_{\rm min}$.  The external
radius $R_{\rm ext}$ of Draco's halo is where it blends smoothly
into the Milky Way halo.  If we define the cuspy halo density as
\begin{equation}
n_{\chi} = A r^{-\gamma}\,,
\label{siseq}
\end{equation}
then $R_{\rm ext}$ and the coefficient $A$ can be jointly set by
equating the $\chi$ density at the edge of Draco's halo to the ambient
Milky Way halo density at Draco's location, $n_{\rm halo}$.  Additionally,
integrating eq.~(\ref{siseq}) out to $R_{\rm ext}$ must give Draco's
total mass.  The following are numerical values for these quantities,
taking $m_{\chi} = 100$~GeV and $\langle \sigma v \rangle_{\chi \bar{\chi}}
= 10^{-26}~{\rm cm}^3/{\rm s}$, listed here for the $\gamma =$ 2 case
(the generalization to $\gamma < 2$ is straightforward):
\begin{eqnarray}
R_{\rm ext} & = & 4.6~{\rm kpc} \nonumber \\
R_{\rm min} & = & 9.4 \times 10^{14}~{\rm cm}           \nonumber \\
n_{\rm halo} & = & 2.2 \times 10^{-5}~{\rm cm}^{-3}     \nonumber \\
n(R_{\rm min}) & = & 5.7 \times 10^9~{\rm cm}^{-3} \,.  \nonumber
\label{numvals}
\end{eqnarray}

As will be discussed in subsection 3.1, the gamma ray flux observed from
a DM clump depends on $R_{\rm min}$, so it is worth considering
whether it is appropriate to use the free fall time
as the cusp formation time.  Since violent relaxation is a complex
process one might imagine some longer characteristic time scale.
Without any particularly compelling choice for this time scale,
we can set a conservative upper limit by using the age of the
universe.  For Draco, using $\tau_{\rm univ} = 15$~Gyr, the resulting
gamma ray flux is found to be about 5 times weaker than it would
be with eq.~(\ref{Rmin}), for any choice of $m_{\chi}$.

The NFW and Moore halo profiles mentioned previously
are the most popularly quoted ``universal profiles'' obtained
from simulations.  Both employ power-law cusps internal to some
scale radius, outside which the density falls off as $r^{-3}$.
The behavior is nearly isothermal in the vicinity of the
scale radius $r_s$.

In calculating $R_{\rm ext}$, we have used the customary NFW
profile for our own Galaxy,
\begin{equation}
n_{\rm halo} = n_0 \left( {r \over r_s} \right)^{-1}
\left[ 1 + {r \over r_s} \right]^{-2}\,,
\label{nfw_halo}
\end{equation}
whose scale radius $r_s \simeq 30$~kpc, total size $R_{\rm ext} \simeq
300$~kpc, and characteristic density $n_0 \simeq 7.5 \times
10^{-4}~{\rm cm}^{-3}$, can be set by the boundary
conditions imposed by the halo density local to our
solar system ($6.5 \times 10^{-25}~{\rm g/cm}^3$), and the total mass of
the Galaxy ($10^{12}~M_{\odot}$), subject to an estimate that
$R_{\rm ext} / r_s \simeq 10$.  As it turns out,
various Galactic halo models roughly agree on the local dark matter
density at Draco's distance from the center, so the choice of NFW
as the Milky Way DM profile is not particularly important here.

The Moore profile~\citep{moore99} is more recent, deriving from
higher resolution computations than NFW.  We therefore include it
in our calculations of Draco.  The Moore profile is
\begin{equation}
n(r) = n_0 \left( {r \over r_s} \right)^{-1.5}
\left[ 1 + \left( {r \over r_s} \right)^{1.5} \right]^{-1} \,,
\label{moore}
\end{equation}
where $n_0 \simeq 2.5 \times 10^{-2}~{\rm cm}^{-3}$ and
$r_s \simeq 0.35$~kpc can be fixed in similar fashion,
using Draco's total mass, the Milky Way density
local to Draco, and a concentration factor
$R_{\rm ext} / r_s \simeq 10$ as boundary conditions.

In a recent paper~\citep{power02}, a convergence study is
described which attempts to resolve central halo densities in
CDM simulations.  The result for a galaxy-sized halo appears to be
a decreasing exponent moving inward from the virial radius
(comparable to $R_{\rm ext}$ in this work), reaching
$\gamma \leq 1.2$, without any particular inner power law
slope convergence.  Because this is a state-of-the-art
simulation, the fact that the inner slope is not especially steep
should not be ignored.  However, the authors note that their
innermost resolved point is only $0.5\%$ of the virial radius,
which is outside our region of interest for Draco, and that the
effect of poor resolution is usually to generate artificially low
central densities.  The reasons which cause this simulation to differ
in inner slope from others like Moore are currently unclear,
but external to a few times the innermost resolved radius, the
various CDM halo profiles are difficult to distinguish.
In any case, the central drop in $\gamma$ is not evident in Draco.

\subsection{Modeling Draco}

The stellar motion measurements from Draco are best fit by
$\gamma \simeq 1.7$ over $\sim$1~kpc~\citep{kweg01}; strong
velocity anisotropies and strong deviations from isothermality are ruled
out over this radial scale.  (Star counts and luminosity density
are insufficient for determining the mass profile, because according
to the same stellar velocty study, mass does not follow light in Draco.)
In the central $\sim$0.2~kpc, $\gamma$ appears to be $\geq 2$, although
other cusp slopes and cores cannot be ruled out at small radii
with the current data set.  So although the observations support
the SIS at intermediate radii, they don't directly speak for radii
smaller than about 10 pc, inside which our annihilation signal
originates.

Since annihilation rate goes as the square of the density of $\chi$
particles ($n_{\chi}^2 \langle \sigma v \rangle_{\chi
\bar{\chi}}~{\rm cm^{-3}\,s^{-1}}$), the choice of $\gamma$ is very
important to the result.  Observed $\gamma$ values from rotation
curves are neither accurate enough nor consistent enough to use uniformly
for this purpose.  Although a positive detection of gamma or radio
signals could identify a particle dark matter source, the difficulty
in choosing $\gamma$ with any confidence has made it hard to
realistically constrain WIMP parameters such as mass and cross
section on the merit of observational upper limits alone.

However, this situation may soon change.  Draco's mass has recently
been shown to be completely dominated by dark matter, with stellar
velocities that favor nearly isothermal organization, and the authors of
these observations contend that the prospects are excellent for improving
the velocity curve data further by sampling more stars~\citep{kweg01}.
This may become a particularly fruitful approach to further particle
DM studies.  So to agree well with current data on Draco, we use halo
models with $\gamma =$ 2.0, 1.9, 1.8, and 1.7, the Moore profile,
and isothermal profiles truncated with cores at 1~pc and 0.1~pc.

Other galaxies currently have more detailed velocity dispersion
or rotation data.  We choose Draco for this study partly for
its proximity, and partly for its lack of features which would be
expected to disrupt the cusp.  Baryonic matter and central
black holes are factors which contaminate this kind of analysis
in other galaxies including our own, but Draco is evidently
little more than a spherical clump of self-gravitating dark matter.
We can't assume that Draco has a steep cusp with complete impunity
because many things can alter the inner structure of the galaxy,
but under a certain set of conditions, such a profile becomes
completely reasonable.  In the next section, we discuss ways in
which a steep cusp might be softened (provided that the relaxation
process progressed far enough to form one initially), and why Draco
should avoid this softening.

\subsection{Retaining a Steep Cusp}

Calculated as in eq.~(\ref{Rmin}), $R_{\rm min}$ is only on the order
of a few AU ($1~{\rm AU} = 1.5 \times 10^{13}$~cm) to a few
$\times 10^{15}$~cm, depending primarily on the value of
$\langle \sigma v \rangle_{\chi \bar{\chi}}$.  It stands to reason
that external influences such as the tidal force from the Milky Way
should have negligible effect on a central cusp that small; tidal
truncation is in fact observationally ruled out below a radius
of 1~kpc~\citep{kweg_draco}.  But it is important
that we consider any other theoretical motivation for
disrupting cusps down to such small scales as
$R_{\rm min}$.  (The appropriate scale to verify for
the validity of the gamma ray limits derived herein is indeed
$R_{\rm min}$; however, as described in subsection 3.2, the radio
synchrotron emanates from within a larger radial size $\sim$~0.01~pc.)
Several such ways to spoil the cusp are addressed here.

(1) It has been proposed that the supermassive black holes found at
the centers of large galaxies will change the dark halo cusps, although
how they will change the cusps is a matter of some debate.  One argument
is that a cusp would steepen into a spike due to accretion of DM onto
the black hole~\citep{gs99,g00}, causing enhanced annihilation
signatures.  If so, then any value of the cusp slope index $\gamma$
over-produces the expected synchrotron signature beyond observational
limits.  Alternatively, if nuclear black holes pair up in a binary
system during a galaxy merger, as one would expect after dynamical
friction pulls both holes to the center of the remnant but before
they merge into one hole, then the two-body interactions of the
holes would throw other masses out of the
center~\citep{mmrb01,mc01}.  In that case, both baryonic and dark
matter would be affected, and the central cusp would be softened;
this result is corroborated by an HST survey~\citep{rhf01}, in that
central mass deficits in galaxies (i.e., the departure from a cusp)
correlate with the masses of their central black holes.

The relevance of this issue depends on whether or not the Draco dwarf
harbors a central black hole.  As a general trend, it appears that
ellipticals and spheroidals brighter than approximately
$2 \times 10^{10}~L_{\odot}$ do have supermassive black holes, and
dimmer galaxies may not~\citep{sg}, although there are exceptions.
For Draco, unless its history is devoid of mergers, the observed
velocity curve should reveal a softened core if there were
a supermassive central hole.  In the absence of direct evidence
for such a hole in Draco, we assume that Draco has none, so
that the concerns of the previous paragraph only afflict other
galaxies.

(2) Could the cusp be scattered by baryonic matter?  \citet{esh01}
have proposed that the missing cusps in galaxy cores are due to
interactions between dark matter and clumpy baryons, in a
scheme which always transfers energy from baryons to WIMPs,
causing the baryonic matter domination at small radii.  Interestingly,
many dwarf galaxies exhibit rising rotation curves out to scales
like $\sim$10~kpc (see the discussion of dwarf galaxies in~\citet{nfw}).
But Draco, with a much higher $M/L$ ratio, has constant velocity across
all observed radii.  So even if baryonic matter
is to blame for destroying cusps in other galaxies, it has a
diminished role in Draco.

(3) Some nonlinear dynamical process among clumpy
dark matter might expand the central region, by way of the halo
forming from a swarm of smaller and denser subhalos which
formed earlier.  For our purposes, the effect is difficult
to estimate, because current simulations have not resolved
much below Draco mass objects.  However, there is no indication
of a cutoff immediately below this mass scale in the
simulations~\citep{mcsqlgg01}.  \citet{m01} finds that even very
small objects ($> 10^3~M_{\odot}$, should they exist) are dense
enough to survive tidal disruption.  But these objects are
too light to experience significant dynamical friction,
so they still behave as collisionless DM, and probably do not
disrupt the smooth halo.  Objects near the mass of the host halo
will settle to the center due to dynamical friction~\citep{mm01},
and be tidally stripped when they pass inside the central
region~\citep{m01}.  So is there an intermediate scale dark object
which does ruin the $\sim 10^8~M_{\odot}$ halo's cusp?  There
is no immediately obvious reason why not.

However, it is interesting to note that globular clusters
(typically $\sim 10^6~M_{\odot}$) are not observed to contain
significant amounts of dark matter (a typical globular cluster ratio of
$2~M_{\odot}/L_{\odot}$ is given by~\citet{bt87}), which may mean that
dark matter clumps of mass scales below those of dwarf galaxies
are suppressed for some reason.  In fact, it has been proposed
that globular clusters could be disrupted remains of larger dwarf
galaxies~\citep{cwm01}, because their numbers are appropriate to
explain the missing dwarf galaxies expected by CDM theory.

The possibility that intermediate mass dark objects don't exist in
large numbers can be dramatically strengthened by upcoming observations.
If substructure DM clumps form on small mass scales like $10^6~M_{\odot}$,
then thousands of such clumps will be observable above the
cosmic microwave background by their annihilation products
(e.g., by the Planck experiment), even if they turn out to
obey gently cusped NFW profiles~\citep{bot02}.
These observations are sensitive to a lower mass cutoff on
substructure DM clumps.  So if more than $\sim$1000 sources are
not found, then it is difficult to see how a steep cusp could
be inhibited in a dwarf galaxy like Draco by dark matter alone.  (This
holds unless the WIMPs do not annihilate like neutralinos, but that
result is immediately consistent with non-detections from Draco anyway.)
If these CMB foreground sources are found, then the results in
section 4 of this paper can be construed as either limits on
neutralino properties or evidence for important dark substructure
dynamics preventing steep cusps in the cores of dwarf galaxies.

\section{Annihilation Signals}

\subsection{Gamma rays}

Supposing that Draco has a dark SIS profile populated by annihilating
neutralinos, one can calculate the gamma ray emission caused by decaying
pions made in the $\chi \bar{\chi}$ annihilation:
$\pi^0 \rightarrow \gamma \gamma$.  The rate of gamma ray
production above a threshold energy $E_0$ per time per volume is
\begin{equation}
q_{\gamma} = n_{\chi}^2 \langle \sigma v \rangle_{\chi \bar{\chi}}
N_{\gamma}(E_{\gamma} > E_0)\,.
\label{qgamma}
\end{equation}
In this work, we adopt a method of approximation that has been used
previously in this context~\citep{bgz92,beu01} for the number of
gamma rays produced above threshold per annihilation,
$N_{\gamma}(E_{\gamma} > E_0)$.  The $\pi^0$ production by
$\chi \bar{\chi}$ annihilation follows
\begin{equation}
{dN_{\pi} \over dx} \simeq K_{\pi} e^{-8 x} / (x^{1.5} + 0.00014)\,,
\label{pi0}
\end{equation}
with $x \equiv E_{\pi}/m_{\chi} c^2$, and $K_{\pi}$ constant.  For
the two photon decay process, the probability of making a photon
per range of energy $E_{\gamma}$ is $2/E_{\pi}$.  Then we have
\begin{equation}
N_{\gamma}(E_{\gamma} > E_0) = K_{\gamma}
\int_{E_0 / m_{\chi} c^2}^1
{ e^{-8 x} \over (x^{1.5} + 0.00014)}\,{dx \over x}\,,
\label{Ngamma}
\end{equation}
where the constant $K_{\gamma}$ is set by requiring that one third of
the total energy released per neutralino annihilation go into gamma
rays, because about one third of the particles produced by $\chi
\bar{\chi}$ are neutral pions, which decay via the two photon
channel.  The remaining two thirds are divided equally among $\pi^+$
and $\pi^-$ particles, whose decay products will be the topic of
the next subsection.  \citet{bgz92} have shown that gamma ray line
flux due to $\chi \bar{\chi} \rightarrow \gamma \gamma$ makes a very
minor correction to the flux predicted above, and we will ignore
it here.

To get the observable gamma ray flux (photons cm$^{-2}$ s$^{-1}$),
integrate eq.~(\ref{qgamma}) over the volume of the source, and
divide by $4 \pi d^2$.  This procedure is specific to the
$r$-dependence of the halo model used, but for example, with an
SIS halo where $n_{\chi} = A r^{-2}$,
\begin{equation}
F_{\gamma}(E_{\gamma} > E_0) = {4 A^2 \over 3 d^2}
\langle \sigma v \rangle_{\chi \bar{\chi}}
N_{\gamma}(E_{\gamma} > E_0) {1 \over R_{\rm min}}\,,
\label{Inugamma}
\end{equation}
where $d$ is the distance from the Earth.  For cusps steeper than
$\gamma = 1.5$, the general result is that the gamma ray flux varies
as $R_{\rm min}^{3 - 2 \gamma}$.  At $\gamma \geq 1.5$, as in the Moore
profile where $\gamma = 1.5$, the integral over the volume of the
source becomes dependent on an outer radius instead of $R_{\rm min}$.
For example, in the Moore profile, the emission region is of
size $\sim r_s$ and the flux depends only very weakly on
$R_{\rm min}$.  For an isothermal halo with a constant density
core, eq.~(\ref{Inugamma}) applies, with $R_{\rm min}$ replaced
by $R_{\rm core}$.  Section 4 gives the results of
this process for Draco.

\subsection{Radio Synchrotron}

The charged pions produced in $\chi \bar{\chi}$ annihilations
decay as
\begin{equation}
\pi^+ \rightarrow \mu^+ \nu_{\mu} \,\, {\rm and} \,\,
\pi^- \rightarrow \mu^- \bar{\nu}_{\mu} \,.
\label{pidecay}
\end{equation}
Muons subsequently decay via 
\begin{equation}
\mu^+ \rightarrow e^+ \bar{\nu}_{\mu} \nu_e \,\, {\rm and} \,\,
\mu^- \rightarrow e^- \nu_{\mu} \bar{\nu}_e \,.
\label{mudecay}
\end{equation}
Electrons and positrons produced in this way will generate synchrotron
radiation if there are ambient magnetic fields.  This calculation was
first done by~\citet{bgz92} for the Galactic center, and is repeated
in~\citet{bot02} and here, with multiple new corrections applied.
(For some parts of this section, cgs units have been provided for
clarity.)

We need to know the number spectrum of electrons and positrons produced
in each neutralino annihilation, $dN_e/dE_e$.  We use a formulaic
simplification of this function~\citep{hill83}, which reduces to
$dN_e/dE_e \sim E_e^{-3/2}$ at low energy, and drops toward zero
as it nears the cutoff $E_e \simeq m_{\chi}$.  The full
expression used is
\begin{equation}
{dN_e \over dE_e} = \int_{E_e}^{m_{\chi} c^2}
\int_{E_{\mu}}^{E_{\mu}/\bar{r}} W_{\pi}
{dN_{\mu}^{(\pi)} \over dE_{\mu}}
{dN_e^{(\mu)} \over dE_e}\,dE_{\pi}\,dE_{\mu}\,,
\label{fragtot}
\end{equation}
where $\bar{r} \equiv (m_{\mu} / m_{\pi})^2$,
\begin{equation}
W_{\pi} = {15 \over 16}
\left( {E_{\pi} \over m_{\chi} c^2} \right)^{-3/2}
\left( 1 - {E_{\pi} \over m_{\chi} c^2} \right)^2\,,
\label{Wpi}
\end{equation}
and
\begin{equation}
{dN_{\mu}^{(\pi)} \over dE_{\mu}} =
{1 \over E_{\pi}}
{m_{\pi}^2 \over m_{\pi}^2 - m_{\mu}^2}
\label{fragpi}
\end{equation}
and
\begin{equation}
{dN_e^{(\mu)} \over dE_e} = {2 \over E_{\mu}}
\left[ {5 \over 6} - {3 \over 2} \left( {E_e \over E_{\mu}}
\right)^2 + {2 \over 3} \left( {E_e \over E_{\mu}} \right)^3
\right]\,.
\label{fragmu}
\end{equation}
These last two equations, eqs.~(\ref{fragpi}) and (\ref{fragmu}),
give the decay products from charged pion and muon decays,
respectively.  Note that $dN_e/dE_e$ without any superscripting
indicates the number spectrum of electrons from the entire chain
of decays following a single $\chi \bar{\chi}$ annihilation.

The charged particle injection is
\begin{equation}
q_e = n_{\chi}^2 \langle \sigma v \rangle_{\chi {\bar \chi}}
\left(\frac{dN_e}{dE_e}\right) \,,
\label{qe}
\end{equation}
so that $q_e \sim E_e^{-3/2} r^{-2 \gamma}$ because
$n_{\chi} \sim r^{-\gamma}$ for a cusp with logarithmic slope $-\gamma$.
Throughout this subsection we will continue to use this generic cusp.
The electron distribution is
\begin{equation}
\frac{dn_e}{dE_e} = q_e \tau \,,
\label{dndE}
\end{equation}
where $\tau = \tau(E_e,r)$ is the average lifetime of an electron of
energy $E_e$.  There are four processes limiting the life of an
electron considered in this work, and for any given combination of
$E_e$ and $r$, the lifetime is the fastest of these processes.
They are:  (1) loss of energy by synchrotron radiation, (2) loss
of energy by inverse Compton scattering (ICS) against the cosmic
microwave background (CMB), (3) loss of energy by ICS against the local
synchrotron photons, and (4) annihilation with positrons.  For pair
annihilation, there is a characteristic time scale; for the other
processes, there is a loss rate $dE_e/dt$ for which
\begin{equation}
\tau \simeq {E_e \over dE_e/dt}\,.
\label{lifetime}
\end{equation}

At a given $E_e$, we find that $e^+ e^-$ pair annihilation can be dominant
in the center of the clump, followed outward by a shell dominated
by ICS against synchrotron photons, followed by the outermost shell
dominated by synchrotron losses and ICS against the CMB.  The
region dominated by ICS against synchrotron photons generates
most of the total flux from the clump.  Synchrotron ICS and
$e^{\pm}$ pair annihilation are discussed below; both have the
effect of removing emitters preferentially from the dense central
regions, reducing the synchrotron signal.

Employing a formula from~\citet{rl}, the synchrotron power is
\begin{equation}
{dE_{{\rm syn}} \over dt} = 1.6 \times 10^{-15} B_{\mu}^2 E_e^2
\,\,\,{\rm erg/s}\,,
\label{dEdtsyn}
\end{equation}
for $B_{\mu} = | \vec{B} |$ in microgauss.  It is convenient to
approximate the synchrotron as if all the radiation were emitted
at the peak frequency (which is different for each electron):
\begin{equation}
\nu_{{\rm peak}} = 1.5 \times 10^{12} B_{\mu} E_e^2 \,\,\,{\rm Hz}\,.
\label{numax}
\end{equation}

Eqs.~(\ref{qe}), (\ref{dEdtsyn}), and (\ref{numax}) combine to
give the radio signal we seek as follows:
\begin{equation}
j_{\nu} = \frac{dn_e}{dE_e} \frac{dE_e}{d\nu} \frac{dE_{\rm syn}}{dt}
\,\,\, \frac{{\rm erg}}{{\rm cm^3\,s\,Hz}} \,.
\label{jnu}
\end{equation}
The luminosity is the integral of $j_{\nu}$ over the
volume of the DM source.

At this point, some details need to be considered.  The above
picture is correct but not complete; three amendments are
analyzed presently.

(1) {\it Inverse Compton scattering} of relativistic electrons
and positrons against the synchrotron photons created in the
DM clump is the most important correction.  ICS power
is~\citep{rl}
\begin{equation}
{dE_{{\rm ics}} \over dt} = {4 \over 3} \sigma_T c \beta^2
\gamma^2 U_{\rm ph}\,,
\label{dEdtics}
\end{equation}
where $\sigma_T = 6.65 \times 10^{-25}~{\rm cm^2}$ is the Thomson
cross section appropriate in the limit $h \nu \ll m_e c^2$,
$\beta \simeq 1$, and $U_{\rm ph}$ is the energy density in
the photon field.  Therefore, the photon population with the
greatest energy density controls the rate of ICS.  For the CMB,
\begin{equation}
U_{\rm cmb} = a T_{\rm cmb}^4\,,
\label{Ucmb}
\end{equation}
for $a = 7.56 \times 10^{-15}~{\rm erg\,cm^{-3}\,K^{-4}}$
and $T_{\rm cmb} = 2.728~{\rm K}$; whereas for the
synchrotron photon population, one must integrate
$j_{\nu}$ over all possible $e^{\pm}$ energies (from
$m_e c^2$ to $m_{\chi} c^2$) and over all lines of sight.

Unfortunately, this double integration for $U_{\rm syn}$
requires {\it a priori} knowledge of the electron distribution
in energy ($E_e$) and in space ($r$), since $dn_e/dE_e$ is
contained in $j_{\nu}$ via eq.~(\ref{jnu}).  The electron
distribution depends on the rate of ICS losses and depends therefore,
in turn, on $U_{\rm syn}$ (via eqs.~(\ref{dndE}), (\ref{lifetime}),
and (\ref{dEdtics})).  Consequently, our full calculation
of $dn_e/dE_e$ is numerical and iterative; however, some
assumptions and simplifications will get us the broad brush
answer as well.

First, assume that the solution takes the form of separable
power laws which we will solve for:
\begin{equation}
{dn_e \over dE_e} (E_e, r) \sim E_e^{\alpha} r^{\beta}\,.
\label{dndEsim1}
\end{equation}
Then we need a further assumption, that the dominant contribution
to the photon density at some point $r$ comes from the region
inside $r$ only.  One might expect this to hold, since all photons
generated inside $r$ will pass through $r$, but only a small
fraction of those produced outside $r$ will.  This is still an
unjustified assumption which can be checked afterward by trying
alternative values for the exponent $\beta$; the numerical results
do validate the assumption as well.

Let's define a shell of width $\Delta r$ such that a photon
spends $\Delta r/c$ time in that region.  Given these assumptions,
and using $r^{\prime}$ as a radial variable of integration,
\begin{equation}
{dU_{\rm syn} \over dE_e} \sim {1 \over 4 \pi r^2 \Delta r}
\int_{R_{\rm min}}^r E_e^{\alpha} r^{\prime \beta}
{dE_{\rm syn} \over dt} {\Delta r \over c} 4 \pi r^{\prime 2}
dr^{\prime}\,.
\label{Usim}
\end{equation}
(The region inside $R_{\rm min}$ makes its own contribution as
well, but it turns out to be small.)
Combining with eq.~(\ref{dEdtsyn}), we get
$dU_{\rm syn}/dE_e \sim E_e^{\alpha + 2} r^{\beta + 1}$.
Then upon integrating over $E_e$, we get
$U_{\rm syn} \sim r^{\beta + 1}$.

From eq.~(\ref{lifetime}),
\begin{equation}
\tau \sim E_e^{-1} r^{-\beta - 1}\,,
\label{lifesim}
\end{equation}
and from eqs.~(\ref{qe}) and (\ref{dndE}),
\begin{equation}
{dn_e \over dE_e} \sim (E_e^{-3/2} r^{-2 \gamma}) (E_e^{-1}
r^{-\beta - 1})\,.
\label{dndEsim2}
\end{equation}
So, simply equating the exponents in eqs.~(\ref{dndEsim1}) and
(\ref{dndEsim2}), we obtain $\alpha = -5/2$ and
$\beta = -\gamma - 1/2$, or
\begin{equation}
{dn_e \over dE_e} \sim E_e^{-5/2} r^{-\gamma - 1/2}\,.
\label{dndEsim3}
\end{equation}

This holds wherever electron losses are fastest by ICS in
the synchrotron photon field.  In the case of losses being
dominated by synchrotron radiation and ICS against the CMB instead,
the loss mechanism is independent of $r$, so we would have
$dn_e/dE_e \sim E_e^{-5/2} r^{-2 \gamma}$.  Since the two forms of
electron losses have different power laws in $r$ (and the same
power law in $E_e$), there is a transition radius $R_{\rm ics,syn}$
(which is independent of $E_e$) between the two behaviors.

For Draco, $R_{\rm ics,syn} \simeq 0.01$~pc, depending loosely
on the WIMP mass and the ambient magnetic field strength.  Inside
this radius, the electron population is limited by ICS against
synchrotron photons and decreases as $r^{-\gamma - 1/2}$; outside
this radius the electron population is limited by synchrotron losses
and ICS against the CMB, and decreases as $r^{-2 \gamma}$.
Interestingly, $R_{\rm ics,syn}$ becomes the characteristic
observable size of the radio source, and it is resolvable to
interferometers with baselines of order $100$~km.

(2) {\it Pair annihilation} between $e^-$ and $e^+$ particles
would have an important role in determining the flux from a
DM source in the absence of the ICS described above.
But in the presence of ICS losses, pair annihilation
makes little difference to the resulting radio flux.

The pair annihilation process is still important, because it
reduces the population of the lower energy pairs, which would
otherwise upscatter photons frequently enough to spoil
the synchrotron spectrum.  Although in much of the DM source,
most electrons Compton scatter with photons, most photons do not
Compton scatter with electrons because they escape first; it
is these synchrotron photons whose survival we are considering
now.  In the case of Draco, and for the most optically
thick line of sight through its diameter, the optical
depth for a photon against synchrotron self Compton
scattering is of order $10^{-3}$, making it a negligible effect
on the resulting observable spectrum.  But one needs to calculate
the pair annihilation rate to discover that fact.

We can estimate the $e^{\pm}$ lifetime for the
case where electrons annihilate before they lose a
significant fraction of their original energy
via synchrotron or ICS as
\begin{equation}
\tau_{e^+} \simeq \frac{1}{\langle n_{e^-}
\sigma_{e^{\pm}} v_{e^-} \rangle}
\label{tau_annih}
\end{equation}
with $v_e \simeq c$, and the angle brackets indicating an average
over $E_{e^-}$.  (This is equally valid if we interchange all
the plus and minus signs, but we will keep track of signs in order
to distinguish between particles and antiparticles, for clarity.)

With eq.~(\ref{dndE}), we have
\begin{equation}
\frac{dn_{e^+}}{dE_{e^+}} = \frac{q_{e^+}}{\langle n_{e^-}
\sigma_{e^{\pm}} v_{e^-} \rangle}.
\label{ne_eq}
\end{equation}
The term in angle brackets is computed from the equations
in~\citet{rs} (also consistent with~\citet{cb}), particularly
\begin{equation}
\langle \sigma v \rangle_{e^{\pm}} = \int_1^{\infty}
f(\gamma_+) \, \int_1^{\infty} f(\gamma_-) \,
(\overline{\sigma v})_{e^{\pm}} \, d\gamma_- \, d\gamma_+ \,,
\label{sven_nsv}
\end{equation}
where $f$ is a distribution function for $e^{\pm}$ of Lorentz
factor $\gamma_{\pm}$, and $(\overline{\sigma v})_{e^{\pm}}$
is the angle averaged reaction rate per pair given explicitly
in~\citet{rs}; here we quote the asymptotic form, although
the full formula has been used in the numerical work
described in this paper:
\begin{equation}
(\overline{\sigma v})_{e^{\pm}} \simeq {\pi e^2 \over m_e c} \,
{1 \over \gamma_+ \gamma_-} \,
({\rm ln} \, 4 \gamma_+ \gamma_- - 2)\,\,\,\,
(\gamma_+,\gamma_- \gg 1)\,.
\label{sven_sv}
\end{equation}
Eq.~(\ref{sven_nsv}) is adapted
by replacing $f(\gamma_-) \rightarrow dn_e/dE_e$, and
$f(\gamma_+) \rightarrow \delta(\gamma_+)$; that is, for a
{\it chosen} positron of energy $\gamma_+$, its survival time
is a function of the electron distribution.

Then, we have  $\langle n_{e^-} \sigma_{e^{\pm}} v_{e^-} \rangle$
expressed as an energy integral which depends on the $e^{\pm}$
distribution that we're trying to solve for.  Therefore this pair
annihilation calculation must be done numerically.  The dependences
here are nested infinitely, which is to say that $\tau_{e^+}$
depends on $n_{e^-}$ which depends on $\tau_{e^-}$ which
depends on $n_{e^+}$ and so on.  It is most efficient to try
to determine the correct dependences {\it a priori}, and then
integrate accordingly.

So if we neglect the natural logarithm factor built into
$(\overline{\sigma v})_{e^{\pm}}$ from eq.~(\ref{sven_sv}),
because it evolves slowly compared with powers of
$\gamma_+$ and $\gamma_-$, then we can proceed to find the
energy dependence for $\tau_{e^+}$ to within a factor of about
$E_{e^+}^{\pm 0.05}$ in accuracy.

The reaction rate for a given positron is
$\langle n_{e^-} \sigma_{e^{\pm}} v_{e^-} \rangle$, which varies
inversely with electron energy (any other hypothesis leads
to a contradiction).  Therefore, the positron will
preferentially annihilate against an electron at the low
end of the energy distribution, so that in eq.~(\ref{ne_eq}),
$\langle \sigma_{e^{\pm}} v_{e^-} \rangle \sim E_{e^+}^{-1}$,
and $n_{e^-}$ is roughly constant.  That is, a given positron of
any energy sees essentially the same distribution of target
electrons, so to a first approximation, it is sufficient to
replace the electron distribution with a single population
at energy $E_{e^-} \simeq m_e c^2$.  By that rationale,
$\tau_{e^+} \sim E_{e^+}$.

With this last result in hand, it becomes useful to use
\begin{equation}
\tau_{e^-} \simeq \tau_{e^+} 
\left( {E_{e^-} \over E_{e^+}} \right)
\label{tweak}
\end{equation}
when doing the full numerical integral over target electron
energies, to avoid the infinite nesting problem.  At low energies
where $dN_e/dE_e \sim E_e^{-3/2}$, this means that $dn_e/dE_e
\sim E^{-1/2}$ and $n_e \sim E^{1/2}$ (where the subscript $e$
refers to either an electron or a positron).

The distribution in space, rather than energy, is easier to
determine.  In order to satisfy eq.~(\ref{ne_eq}) with the
injection $q_e \sim r^{-2 \gamma}$ set by a cuspy DM profile,
we must have $dn_e/dE_e \sim r^{-\gamma}$.  (As before, $- \gamma$
is the halo profile logarithmic slope, while $\gamma_+$ and
$\gamma_-$ are Lorentz factors.)  Then for a region
whose $e^{\pm}$ losses are fastest by pair annihilation, we have
\begin{equation}
{dn_e \over dE_e} \sim E_e^{-1/2} r^{-\gamma}\,.
\label{dndEsim_annih}
\end{equation}

So now we have another
transition radius between the $r^{-\gamma}$ pair annihilation
distribution and the $r^{-\gamma - 1/2}$ ICS distribution, which we
designate $R_{\rm ann,ics}$.  If $R_{\rm ann,ics} > R_{\rm min}$,
then we add a pair annihilation dominated central region to the
picture described above regarding ICS.  But for most choices
of $m_{\chi}$ and Draco's magnetic field strength,
$R_{\rm ann,ics} < R_{\rm min}$ for electrons radiating
at frequencies of interest, and the distribution flattens
at the center before the density can get high enough
to create a pair annihilation dominated region.

(3) {\it Synchrotron self absorption} (SSA) is where an electron
in a magnetic field can absorb the synchrotron photon's energy.
Still linking photon frequency to the energy of its emitting
electron with eq.~(\ref{numax}) for simplicity, the appropriate
adaptation from~\citet{rl} yields the per-unit-length absorption
coefficient:
\begin{equation}
\alpha_{\nu} = -{c^2 \over 8 \pi \nu} {dE_{{\rm syn}} \over dt}
\left[ {\partial \over \partial \nu}
\left( {1 \over \nu} {dn_e \over d\nu} \right) \right]\,.
\label{alpha_ssa}
\end{equation}
SSA absorbers  of a particular photon are electrons of similar
energy to the original synchrotron emitter which made the photon.
As such, SSA is only effective below some
$\nu_{\rm crit}$, where the corresponding electrons are more
numerous.

The source function for an absorbing source is
$S_{\nu} = j_{\nu} / \alpha_{\nu}$, and the
optical depth is
\begin{equation}
\tau_{\nu} = \int \alpha_{\nu}\,dz\,,
\label{taudef}
\end{equation}
where $z$ is a line-of-sight coordinate.  The flux
density from such a source at distance $d$ from the Earth
is obtained by summing over each line of sight through the
DM clump:
\begin{equation}
I_{\nu} = {1 \over 4 \pi d^2} \int_0^{R_{\rm cl}} 2 \pi 
b S_{\nu}(1-e^{-\tau_{\nu}})\,db\,\,\,
{{\rm erg} \over {\rm cm^2\,s\,Hz}}\,.
\label{Inu_Snu}
\end{equation}

The SSA cutoff frequency $\nu_{\rm crit}$ depends in part on the absorber
density, and without ICS or $e^{\pm}$ pair annihilation effects,
SSA would be the most important correction to the calculated flux in
a steep cusp.  SSA has been included in the calculations performed here,
but since the central absorber population is diminished
by ICS, it is of small importance.  Using appropriate parameters
for Draco, we find $\nu_{\rm crit} \simeq 10$~MHz, whereas
the limiting observations discussed in the next section are at
4.9~GHz.  (It is fortunate that SSA has little impact on the
problem, because otherwise it would create serious complications
in the calculation of ICS against the synchrotron photon field.)

\section{Results}

Supersymmetric dark matter has a number of variable parameters
which affect its annihilation and clustering properties.
However, these parameters boil down to only two relevant
quantities for annihilation signature searches, and those are
$m_{\chi}$ and $\langle \sigma v \rangle_{\chi \bar{\chi}}$.
Given external constraints on a dark matter source (such as
Draco's total mass and the matter density local to it, in the
present case), the mass $m_{\chi}$ determines the $\chi$ number
density.  The more massive the particle, the fewer are needed,
so the annihilations are less frequent.
We present here the collection of $m_{\chi}$ and
$\langle \sigma v \rangle_{\chi \bar{\chi}}$ values available
to the neutralino, derived from observations of Draco.

Before proceeding to these limits, we note
for comparison another indirect detection approach
considered in the literature, that of detecting
neutrinos (as from eqs.~(\ref{pidecay}) and (\ref{mudecay}))
coming from neutralino annihilations at the center of the
Earth and the Sun.  By elastic scattering with nuclei in the
Sun or Earth, a WIMP can lose enough energy to be
gravitationally captured by that body.  This captured
population is the source of the neutrino signal in question.
Supersymmetric parameter space can be ruled out in this way
by neutrino telescopes such as Baksan or AMANDA (for example,
see~\citet{beg97} and~\citet{amanda02}).  This method can
currently eliminate more parameter space than direct detection
experiments, and less than the Draco EGRET limits as calculated
herein.  The constraints from Draco require the extrapolation
of its halo density profile to small radii; the constraints from
neutrino telescopes instead require the additional steps associated
with elastic scattering, capture, and neutrino propagation
through matter.  Of course, this approach and the one used
in the present work are both indirect detection scenarios; for
a review of direct experimental searches, see
e.g.~\citet{morales01}.

\subsection{Gamma Ray Constraints}

Gamma ray data specifically on Draco are lacking, so the current
limiting observations come from an all-sky survey.  The EGRET
(Energetic Gamma Ray Experiment Telescope) instrument, flying on
the CGRO satellite (Compton Gamma Ray Observatory), is a
pair production telescope (see~\citet{egret96} for an
introduction to EGRET and its data analysis).
As such, its useful beam width is around $30^{\rm o}$,
making it well suited for a full sky survey; its energy
range is $\sim$30~MeV to below $\sim$100~GeV~\citep{egret98}.

EGRET sources are usually counted by their photons above
100~MeV; however,~\citet{egret97} have performed a separate
binning of the data, making a catalog of gamma ray sources
above 1~GeV, which is of particular interest here.  The least
significant detection ($4 \sigma$) in the catalog, with the least
flux, is the Large Magellanic Cloud (LMC), at $(1.1 \pm 0.4) \times
10^{-8}$~photons~${\rm cm^{-2} \,\, s^{-1}}$ above 1~GeV.
We therefore take $10^{-8}$~photons~${\rm cm^{-2} \,\, s^{-1}}$ as
the flux limit for detection, and require that Draco emit less.

In fact, since this limit is only based on non-inclusion in the
GeV catalog, specific knowledge of the exposure on Draco could
result in a bound somewhat better that the one derived herein.
A subsequent analysis of the EGRET data provides an
estimate of the instrument's extragalactic gamma ray
background~\citep{egret98}. With Draco located well outside the
Galactic disk ($\mathit{l} = 86.37^{\rm o}, \mathit{b} =
34.71^{\rm o}$~\citep{vdb00}), this extragalactic component is
the primary background source.  The extragalactic flux, adapted
to the GeV catalog's 30~arcmin pixel size,
\begin{equation}
F_{\gamma} (E_{\gamma} > E_0) = 8.6 \times 10^{-11}
\left( {E_0 \over {\rm GeV}} \right)^{-1.1}~{{\rm photons} \over
{\rm cm^2\,\,s}}
\label{Ngam_eg}
\end{equation}
is well below the level of one photon per minimum EGRET exposure.
We therefore separately consider this background in our
calculations, conservatively assuming that EGRET's exposure
on Draco is their minimum exposure (roughly true),
for the case of no photons detected within 30~arcmin.  We present
this as an example of the type of improvement possible without
collecting new data.

\begin{figure}[t]
\vspace*{2.0mm}
\includegraphics[angle=-90,width=8.3cm]{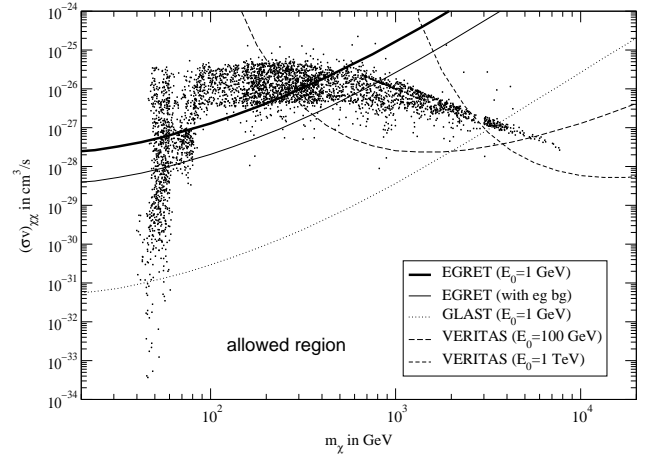}
\caption{The $m_{\chi} - \langle \sigma v \rangle_{\chi
\bar{\chi}}$ plane for neutralino dark matter.  Curves indicate
results where the SIS halo profile has been used.  Shown
as solid lines are the gamma ray constraints from non-inclusion
in the EGRET all-sky survey, where the thick line is the
current bound and the thin line is a potential improvement
using an estimate of the extragalactic gamma ray
background (see text).  Both employ a threshold $E_0 = 1$~GeV.
Also shown are two dashed lines showing upcoming VERITAS
constraints with cutoffs at 100~GeV and 1~TeV, and a
dotted line for GLAST at $E_0 = 1$~GeV
(sensitivities for all of these taken from~\citet{veritas}).
The region below the curves is allowed by gamma
ray observations.  Uncertainties in Draco's mass and distance lead to
at most a factor of 3 change in the vertical axis position of
these curves.  The dots denote typical supersymmetric
dark matter properties~\citep{darksusy}.}
\label{fig_msvgam}
\end{figure}

\begin{figure}[t]
\vspace*{2.0mm}
\includegraphics[angle=-90,width=8.3cm]{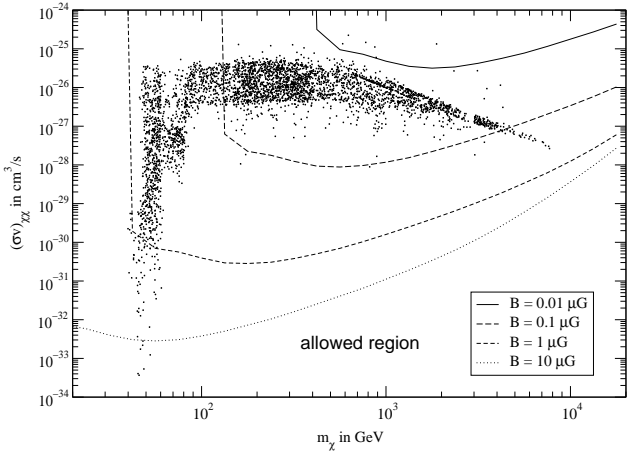}
\caption{The $m_{\chi} - \langle \sigma v \rangle_{\chi
\bar{\chi}}$ plane for neutralino dark matter.  Curves indicate
results where the SIS halo profile has been used.  Shown
as thin lines are the radio synchrotron constraints from
VLA observations.  From top to bottom, these lines
correspond to magnetic fields of 0.01, 0.1, 1.0, and 10.0
$\mu {\rm G}$ in the core of Draco.  The region below these curves
is allowed by radio observations.  Uncertainties in Draco's mass
and distance lead to at most a factor of 3 change in the vertical
axis position of these curves.  The dots denote typical
supersymmetric dark matter properties.}
\label{fig_msv}
\end{figure}

In figure~\ref{fig_msvgam}, we plot the neutralino parameter
space.  In this figure, we have presented limits derived by
use of the SIS halo profile with different curves giving
different present and future observational bounds; a
comparison of halo models will follow in subsection 4.3.
Both EGRET constraints are shown:  the current bound
(thick solid line), and the example bound with the newer
extragalactic background (thin solid line).  Points above the
curves are prohibited by these observational constraints
from Draco.

Two notes should be made regarding accuracy.  First, the curves in the
figure assume that $M_{\rm Draco} = 8.6 \times 10^7~M_{\odot}$, which
is measured using observations of its luminous matter~\citep{kweg01}.
So this mass samples the inner halo, and the total mass could be
higher.  In this sense, the curves plotted should be considered
conservative.  Second, recall from the discussion of $R_{\rm min}$
in subsection 2.1 that a higher estimate of $R_{\rm min}$
would result from a slower relaxation process.  At most, this
would make these curves less constraining (i.e., raise them) by
a factor of about 5.

The dots in the figure survey the possible combinations of
$m_{\chi}$ and $\langle \sigma v \rangle_{\chi \bar{\chi}}$ for
neutralino dark matter~\citep{darksusy}.  The dotted region
is characteristic of supersymmetric parameters often considered
in the literature (e.g.,~\citet{gs99,baltz99}).  They are constrained
by the requirement that the relic density be cosmologically
interesting; that is, large enough to account for a significant
component of the matter density without exceeding it.  The choice
of these ``interesting'' values is somewhat arbitrary and
authors differ slightly, but a restriction like
$0.05 \leq \Omega_{\chi} h^2 \leq 0.5$
as depicted in the figure is typical.
It should be noted that values of $m_{\chi} < 38$~GeV
are disfavored by accelerator experiments~\citep{aleph}.

Future gamma ray telescopes offer better constraints.  The
upcoming satellite mission GLAST (Gamma ray Large Area Space
Telescope), will offer much improved constraints as
shown in the figure~\citep{glast}.  It is expected to launch
in 2005 for a two year all-sky survey, effective between 20~MeV and
300~GeV, with a sensitivity about 20 times better than EGRET at 1~GeV.
As can be seen in the figure, GLAST has the potential to rule out
most of the neutralino DM parameter space.

Apart from satellite missions, atmospheric
Cherenkov telescopes (ACTs) are ideal for measuring at
high energy.  Rather than detecting the photons via
interactions inside the telescope, ACTs count gamma rays by the air
showers they produce in the Earth's atmosphere.  Specifically, they
respond to Cherenkov light produced by charged particles in the shower
traveling faster than the speed of light in air.  The next generation
gamma ray telescope VERITAS~\citep{veritas} can provide significant
benefits for this study with a typical exposure on Draco (we quote
here an exposure of 50 hours).  Curves for VERITAS
with threshold energies of 100~GeV (long dashes) and 1~TeV (short
dashes) are depicted in the figure.  VERITAS is sensitive between
about 50~GeV and 50~TeV; it is expected to see first light in 2002
and become fully operational in 2005.

\subsection{Radio Constraints}

The applicable radio continuum limits for Draco were performed
by~\citet{fg79}, using the VLA (Very Large Array) facility in
New Mexico.  At 4.9~GHz, they report no flux from Draco above
2~mJy, at the $3 \sigma$ level.  Their search spanned a radius
of 4~arcmin around the galaxy center.

Two additional assumptions beyond those discussed in section 2
are needed in order to place a radio limit, because synchrotron
emission requires magnetic fields.

First, we must assume a magnetic field strength $| \vec{B} |$
for the core of Draco.  No such estimates exist in the literature,
so this paper must be viewed as constraining
either the neutralino or Draco's magnetic properties.  To give
some bearing on the estimation of $| \vec{B} |$, we can consider
measurements for other dwarf galaxies.  The Magellanic Clouds, for
example, each carry fields of $\sim$5~$\mu {\rm G}$~\citep{pohl93}.
A survey of low surface brightness dwarf galaxies (similar to
Draco) leads~\citet{klein92} to infer typical field strengths
between 2 and 4~$\mu {\rm G}$, although this survey was conducted
near 5 GHz and Draco specifically gives no measured signal there. 
The dwarf irregular NGC 4449 was found to harbor $\sim$14~$\mu
{\rm G}$ fields despite the lack of ordered rotation~\citep{cbkku00},
suggesting that some non-dynamo process may be sufficient to
generate microgauss fields.  Based on these data, it seems
appropriate to expect $\sim 0.1 - 1~\mu {\rm G}$ in Draco,
although at present there is no compelling evidence.

Second, not only must there exist magnetic fields, but they also
must be sufficiently constricted, so as to trap electrons and
positrons long enough for them to radiate.  Implicit in the
computation of flux from DM sources in this paper is the
assumption that $e^{\pm}$ particles never stray far from their
birth places $-$ an assumption of strong magnetic confinement.
To begin contemplating this, consider that the gyroradius of
any electron relevant to this problem is significantly smaller
than the smallest length scale applicable; that is,
$R_{\rm gyro} \ll R_{\rm min}$.  So if we assume that electrons
are reflected back and forth along a field line every
$10 - 100~R_{\rm gyro}$, then strong confinement is justified.
\citet{klein91} notes that magnetic fields do in fact appear
to be highly disordered in some dwarf galaxies (Draco not
considered), based on lack of radio polarization observed.
The confinement time needs to be longer than the
inverse Compton time scale against the synchrotron photon field
in order to validate our model.  This duration is typically
around $10^7 - 10^8$~years for key emitting regions in Draco's
halo, depending on neutralino parameters, for 4.9~GHz radiation
as limited by the VLA.

These caveats noted, the synchrotron constraints are plotted
in figure~\ref{fig_msv}, where again the points above each curve
can be ruled out by it.  Limits for 0.01, 0.1, 1, and
10~$\mu {\rm G}$ fields are shown here for the SIS halo;
this choice will be varied in section 4.3.  As with
figure~\ref{fig_msvgam}, the possibility that
$M_{\rm Draco}$ was underestimated makes these
curves conservative.

Although the radio limits are less direct because of assumptions
that must be made about the $\vec{B}$ field, they can be
stronger than current gamma ray limits.  For example, if the field
is found to be at least $1~\mu {\rm G}$ and adequately confining,
then $\langle \sigma v \rangle_{\chi \bar{\chi}}$ for a 100~GeV
neutralino is restricted to be below $\sim 4 \times
10^{-31}~{\rm cm^3/s}$, as compared with
$\sim 10^{-28}~{\rm cm^3/s}$ in gamma rays.
Heavier neutralinos are strongly constrained at $0.1~\mu {\rm G}$,
but they aren't limited at all by current gamma measurements.
On the other hand, if the magnetic fields are weak, then
synchrotron offers no constraint on light $\chi$ particles.
For example, a 100~GeV neutralino's annihilation products aren't
energetic enough to radiate at 4.9~GHz (where the VLA limit is)
if the field is $\sim$0.1$~\mu {\rm G}$ or below.

As in figure~\ref{fig_msvgam}, the dots indicate the
SUSY parameter space.  Here we see the importance of the magnetic
field.  At $1~\mu {\rm G}$ one can exclude
the majority of neutralino dark matter candidates; at
$0.01~\mu {\rm G}$ one can exclude almost none of it.

One further point worth noting is that the gamma ray and radio
analyses, taken together, permit us to constrain the
magnetic field properties of Draco.  If a future gamma ray
detection is made, then the corresponding line in the
$m_{\chi} - \langle \sigma v \rangle_{\chi \bar{\chi}}$ plane
can be transposed to figure~\ref{fig_msv}, with the result that
some values of $| \vec{B} |$ will not be allowed.

\subsection{Other Halo Profiles}

Figures~\ref{fig_msvgam} and~\ref{fig_msv} solely employ the SIS
to generate the limiting curves.  In this subsection, we consider the
sensitivity of our results to the choice of halo profile.

\begin{figure}[t]
\vspace*{2.0mm}
\includegraphics[angle=-90,width=8.3cm]{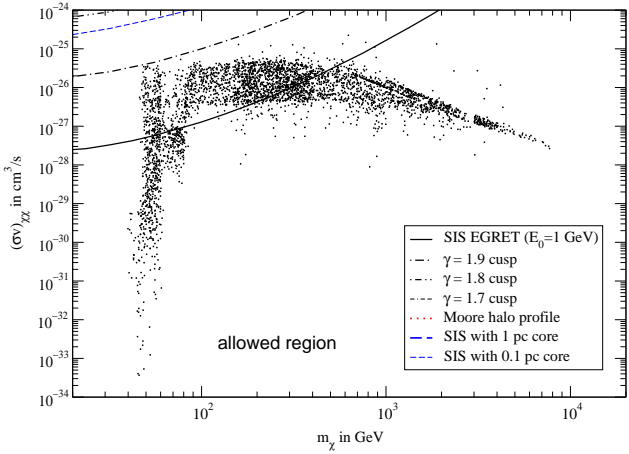}
\caption{Parameter space constraints from the EGRET 1~GeV
catalog with seven different halo models for Draco.  The halos
are:  cusps with $\gamma =$ 2.0 (solid line), 1.9 (dot-dash),
1.8 (dot-dot-dash), and 1.7 (dot-dash-dash), Moore (dotted line),
SIS with constant density core at 1~pc (thick dashed line),
and SIS with constant density core at 0.1~pc (thin dashed line).
The thickest solid line is repeated from the current EGRET bound
given in figure~\ref{fig_msvgam}.  Lines not shown indicate that
the corresponding halo yields no constraints on this plot.}
\label{fig_msvgam_EGRtweak}
\end{figure}

\begin{figure}[t]
\vspace*{2.0mm}
\includegraphics[angle=-90,width=8.3cm]{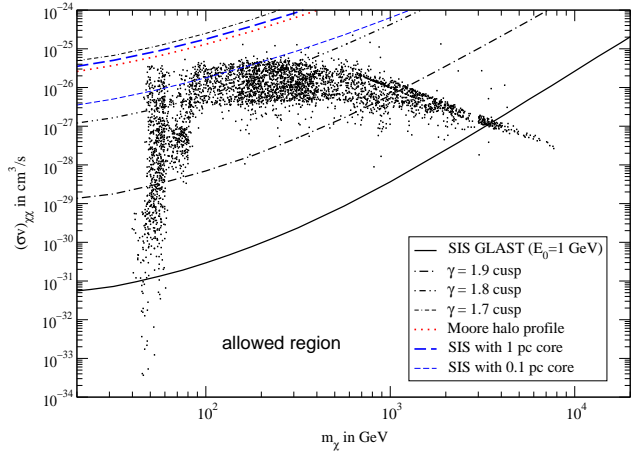}
\caption{Same as figure~\ref{fig_msvgam_EGRtweak}, but now showing
modifications to the GLAST line from figure~\ref{fig_msvgam},
rather than EGRET.}
\label{fig_msvgam_GLAtweak}
\end{figure}

\begin{figure}[t]
\vspace*{2.0mm}
\includegraphics[angle=-90,width=8.3cm]{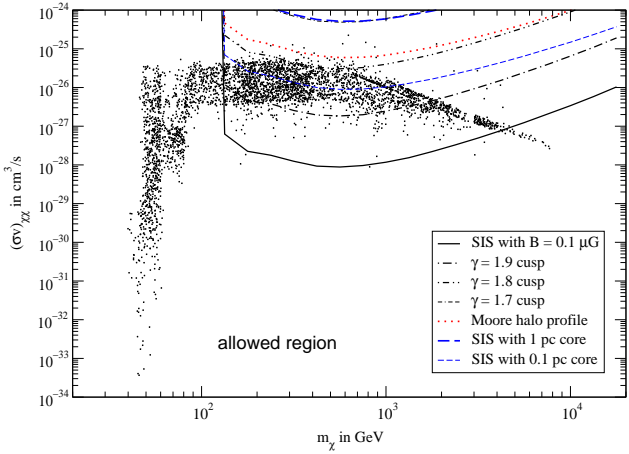}
\caption{Parameter space constraints from the VLA
with seven different halo models for Draco, assuming a magnetic
field strength $| \vec{B} | = 0.1~\mu {\rm G}$.  The halo
modifications are the same as those in
figures~\ref{fig_msvgam_EGRtweak} and \ref{fig_msvgam_GLAtweak}.}
\label{fig_msv.1_tweak}
\end{figure}

\begin{figure}[t]
\vspace*{2.0mm}
\includegraphics[angle=-90,width=8.3cm]{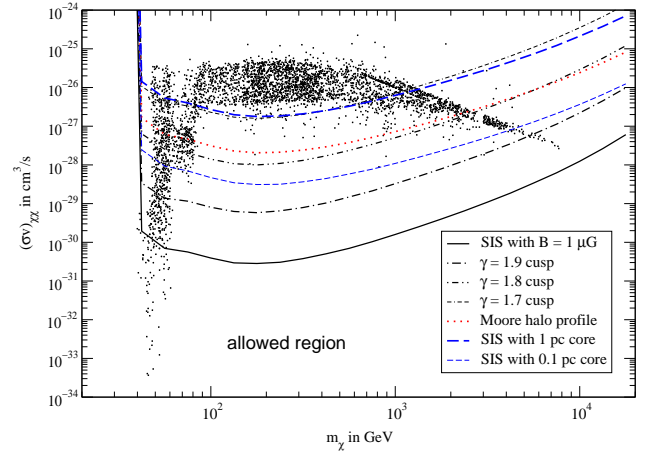}
\caption{Same as figure~\ref{fig_msv.1_tweak}, but now showing
modifications to the $| \vec{B} | = 1~\mu {\rm G}$ line from
figure~\ref{fig_msv}.}
\label{fig_msv1_tweak}
\end{figure}

In figure~\ref{fig_msvgam_EGRtweak}, we present modified gamma
ray results corresponding to the current EGRET bound shown as the
thick line in figure~\ref{fig_msvgam}.  All curves on this plot
correspond to the same EGRET observations above the $E_0 = 1$~GeV
threshold, but for different halo models.
Figure~\ref{fig_msvgam_GLAtweak} presents the same
halos, but for the GLAST curve from figure~\ref{fig_msvgam}
with $E_0 = 1$~GeV.

The halo models considered all assume spherical symmetry.
They are:  cusped profiles with $\gamma =$ 2.0 (SIS), 1.9, 1.8,
and 1.7, the Moore profile as described in subsection 2.1, and
the SIS truncated with constant density cores of radius 1~pc
and 0.1~pc (the velocity dispersion data extend down to
$\sim 10$~pc).  Figures which omit some of these curves do so
because those curves do not constrain any of the parameter space
shown.  For example, with Draco modeled by the Moore profile,
the EGRET data are insufficient to contrain
$\langle \sigma v \rangle_{\chi \bar{\chi}}$
below $\sim 10^{-24}~{\rm cm^3/s}$; one can do somewhat better
in the future with GLAST (figure~\ref{fig_msvgam_GLAtweak}).

For the radio case, the corresponding results are presented in
figure~\ref{fig_msv.1_tweak} for $| \vec{B} | = 0.1~\mu {\rm G}$,
and in figure~\ref{fig_msv1_tweak} for $| \vec{B} | = 1~\mu
{\rm G}$.  In each figure, the same seven halo models are
shown as in figures~\ref{fig_msvgam_EGRtweak} and
\ref{fig_msvgam_GLAtweak}, including the appropriate SIS curve from
figure~\ref{fig_msv}.

The cusps $\gamma =$ 2.0, 1.9, 1.8, and 1.7 are calculated
with the full inverse Compton scattering corrections described in
subsection 3.2.  Applying eq.~(\ref{dndEsim3}) yields $dn_e/dE_e$
dependences which go as $r^{-2.5}$, $r^{-2.4}$, $r^{-2.3}$,
and $r^{-2.2}$, respectively.

Inverse Compton scattering does not play a significant role in
the SIS core cases or in the Moore case, because in each of these
halos, the primary emitting region is outside of $R_{\rm ics,syn}$.
The Moore case turns out to create a diffuse enough collection
of electrons and positrons that synchrotron self absorption
is also negligible, although SSA does significantly affect the
results for the isothermal core models.

It is interesting to note that for both gamma ray and radio
constraints, the Moore profile yields more flux than a $\gamma = 1.7$
cusp, even though the inner Moore halo only has $\gamma = 1.5$.  This
is a result of the fact that the Moore profile does not extend
as $r^{-1.5}$ indefinitely, but rather is concentrated into $r_s$.
This concentration effect makes up for the lower $\gamma$ exponent.

From figures~\ref{fig_msvgam_EGRtweak} through~\ref{fig_msv1_tweak},
it is clear that the choice of halo model is of great importance in
determining the resulting dark matter constraints.  Even
$\rho \sim r^{-1.9}$ yields weak limits with EGRET data.  With
GLAST, many of the steeper models are valuable, but less dense ones
are not.  In the radio, a magnetic field of $0.1~\mu {\rm G}$
is roughly as constraining at high $m_{\chi}$ values as GLAST is
at lower $m_{\chi}$ values.  At $1~\mu {\rm G}$, the situation is
much better, with all examined profiles generating strong constraints.
In this case, the Moore halo too constrains WIMPs well, although
the Moore profile is not particularly constraining in any
of the other cases considered.  The primary lesson from these
example cases is that the results do vary appreciably even
among relatively steep cusps, underscoring the importance of
pinning down Draco's central density structure accurately.

\section{Summary}

This paper applies gamma ray and radio observational limits to
the behavior of dark matter in the Draco dwarf galaxy, only
79~kpc distant from the Earth.  Recent measurements of stellar
velocities in Draco suggest that it is extremely dark matter
dominated, and that its halo distribution is
nearly isothermal.  Because such a halo should be
densely concentrated at the center, we expect WIMP annihilation
to proceed rapidly there, producing observable gamma ray
and synchrotron radiation.

Current limits derived herein impose
restrictions on the nature of the dark matter particle.  If it
is a neutralino (or simply annihilates into pions like one) then
figures~\ref{fig_msvgam} through~\ref{fig_msv1_tweak}
constrain the mass and annihilation cross section
available to it.  These results emphasize the importance of
the choice of halo model for Draco.  Our strongest constraints
emerge with the SIS profile.  In this case, EGRET data restrict
much of the supersymmetric parameter space, and if Draco has a
suitable magnetic field, then the VLA data can exclude significantly
more of it.  Other halo profiles with a core or a less steep cusp
generally become significantly constraining with future gamma ray
experiments, or in the radio with magnetic fields in the
$\mu {\rm G}$ range.

An important conclusion from this work is that
further particle dark matter research would be
well served by increasingly detailed studies of Draco.

\begin{acknowledgements}

Thanks to Angela Olinto, Josh Frieman, Rich Kron, and Jim Truran
for evaluating the topic, suggesting avenues to consider, and
examining the manuscript.  I also thank Pasquale Blasi for
introducing me to this line of research, Paolo Gondolo for
providing SUSY model points, and Argyro Tasitsiomi for many
useful and relevant conversations.  This work was supported
by the NSF through grant AST-0071235 and DOE grant
DE-FG0291 ER40606.

\end{acknowledgements}

\end{document}